\documentclass[12pt]{iopart}
\usepackage{epsf}
\begin{document}

\title[Spectroscopy of Dark Soliton States]
       {Spectroscopy of Dark Soliton States in Bose-Einstein Condensates}

\author{K. Bongs\dag\ , S. Burger\ddag\, D. Hellweg\S\, M. Kottke\S\, S. Dettmer\S\, T. Rinkleff\S\,
        L. Cacciapuoti\S\, J. Arlt\S\, K. Sengstock\dag\ and W. Ertmer\S }

\address{\dag\ Institut f\"ur Laserphysik, Universit\"at Hamburg, Jungiusstr. 9, 20355 Hamburg, Germany}

\address{\ddag\ LENS, Dipartimento di Fisica, Universit\`{a} di Firenze and Istituto Nazionale per la
         Fisica della Materia, Largo E. Fermi 2, I-50125 Firenze,
         Italy}

\address{\S\ Institut f\"ur Quantenoptik, Universit\"at Hannover, Welfengarten1, 30167 Hannover, Germany}

\ead{kai.bongs@physnet.uni-hamburg.de}

\begin{abstract}
Experimental and numerical studies of the velocity field of dark
solitons in {B}ose-{E}instein condensates are presented. The
formation process after phase imprinting as well as the
propagation of the emerging soliton are investigated using
spatially resolved Bragg-spectroscopy of soliton states in
Bose-Einstein condensates of $^{87}$Rubidium. A comparison of
experimental data to results from numerical simulations of the
Gross-Pitaevskii equation clearly identifies the flux underlying a
dark soliton propagating in a Bose-Einstein condensate. The
results allow further optimization of the phase imprinting method
for creating collective exitations of Bose-Einstein condensates.
\end{abstract}


\submitto{\JOB}


\maketitle

\section{Introduction}
With the realization of {B}ose-{E}instein condensation (BEC) in
dilute atomic gases~\cite{Anderson1995a,Davis1995b,Bradley1995a},
a well controllable and easily observable macroscopic wavefunction
has become accessible for a wide variety of fundamental
experiments. Since the early experiments on oscillations of the
condensate~\cite{Jin1996a,Mewes1996b} and
phonons~\cite{Andrews1997a}, excitations of the condensate
wavefunction have been subject to intense investigation.
Signatures of superfluidity in Bose-Einstein condensates were
first observed in the suppression of excitations below a critical
excitation velocity \cite{Raman1999a,Onofrio2000b} and in the
analysis of scissors mode excitations~\cite{Marago2000a}. Recently
another striking aspect of superfluidity in BEC has been studied
by exciting vortices in two-component
condensates~\cite{Matthews1999a} and vortices and vortex arrays in
stirred
condensates~\cite{Madison2000a,Chevy2000a,Hodby2001a,Inouye2001a,Abo-Shaeer2001a,Haljan2001b}.
Vortices as topological excitations of the condensate wavefunction
represent particularly interesting phenomena and are still a very
active area of research. Dark solitons in Bose-Einstein
condensates are in several respects the one-dimensional
counterpart to vortices. While the phase of the vortex
wavefunction winds up around a zero density 'core', in a dark
soliton state regions of constant phase are separated by a minimum
in density and a steep phase gradient. Dark solitons are
fundamental excitations in one dimensional systems described by
nonlinear wave equations and thus a solution of the Gross
Pitaevskii equation, which describes the condensate wave function
\cite{Pitaevskii1961a}. In Bose-Einstein condensates, the creation
of dark solitons using a phase imprinting
method~\cite{Dobrek1999a} has been investigated in a strongly
elongated geometry in BECs of $^{87}$Rb~\cite{Burger1999a} and in
nearly spherical BECs of $^{23}$Na~\cite{Denschlag2000a}. Recently
bright solitons in Bose-Einstein condensates with attractive
interactions \cite{Khaykovich2002a,Strecker2002a} and filled
solitons \cite{Anderson2001a} have also been observed .

The experiments on dark solitons have so far concentrated on the
temporal evolution of the density distribution of the
Bose-Einstein condensate, which showed the motion and decay of
these excitations. In this article we extend these investigations
to the velocity field which underlies a dark soliton state in a
Bose-Einstein condensate.

\section{Properties of dark solitons}
In general Solitons are fundamental wave-packet-like excitations
of one dimensional nonlinear systems. One characteristic of them
is that they do not disperse but preserve their shape due to a
balance between dispersive and stabilizing effects in these
systems. Dark solitons in Bose-Einstein condensates form density
minima, which self stabilize due to a balance of the repulsive
mean field interaction which would be minimized by compressing the
minimum and quantum pressure. The quantum pressure is related to
the kinetic energy associated with the dark soliton velocity
field, which is minimized by widening the minimum.

More formally a dark soliton in a homogeneous Bose-Einstein
condensate of density $n_0$ is described by the one dimensional
wavefunction \cite{Fedichev1999a}
\begin{equation}
\Psi_k(x) = \sqrt{n_0}\left( i\frac{v_k}{c_s}+
\sqrt{1-\frac{v_k^2}{c_s^2}}
\tanh\left[\frac{x-x_k}{l_0}\sqrt{1-\frac{v_k^2}{c_s^2}}\right]\right)
, \label{soliton_eqn}
\end{equation}
with the position $x_k$ and velocity $v_k$ of the dark soliton,
the correlation length $l_0=(4\pi a n_0)^{-1/2}$, and the speed of
sound $c_s=\sqrt{4\pi a n_0}\hbar/m$, where $m$ is the atom mass.
Note that a so called black soliton with zero density at the
minimum and a $\pi $ phase jump is obtained for $v_k=0$, while for
all other cases the condensate wave function has nonzero amplitude
at the position of the density minimum which now moves along the
condensate and is associated with a finite phase gradient.

\begin{figure}
\begin{center}
\epsfxsize=3in \epsfbox{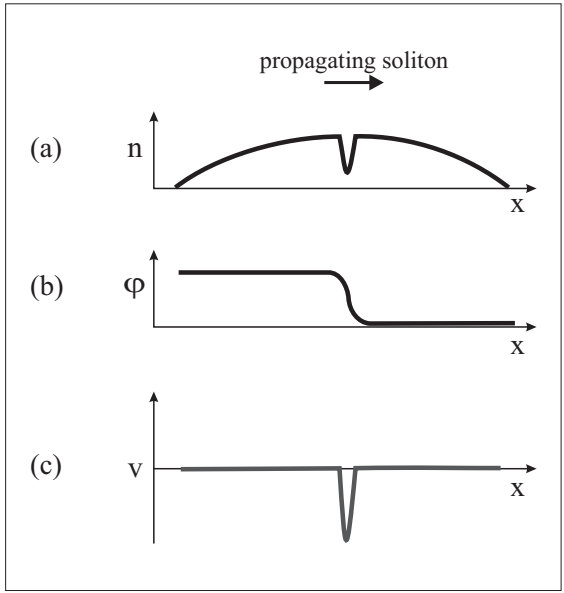}
\end{center}
\vspace{0in} \caption {\small Schematic of the density
distribution (a), phase distribution (b), and the resulting
velocity distribution (c) of a dark soliton propagating in a BEC.
} \label{fig1}
\end{figure}

Consequently dark solitons in Bose-Einstein condensates are
density minima associated with a steep phase gradient of the
condensate wavefunction. This gradient of the phase distribution
$\Phi(\vec{r}\,)$ of the macroscopic Bose-Einstein condensate
wavefunction defines a probability current density, which can be
interpreted as a superfluid velocity field
$v(\vec{r}\,)=\frac{\hbar}{m}\nabla\Phi(\vec{r}\,)$. Thus, the
steep gradient of the phase at the dark soliton leads to a
localized peak in the velocity distribution of the BEC. The
corresponding BEC density, phase and velocity distributions are
shown schematically in Fig. \ref{fig1}.

Equation \ref{soliton_eqn} is strictly valid only for infinite
homogeneous one-dimensional systems but many qualitative features
remain valid in more realistic experimental cases. However the
effect of the trapping potential, the formation process and the
dynamics during time-of-flight expansion are not included. In
order to model these cases more accurately, we numerically solve
the Gross-Pitaevski equation for the case of inhomogenous
condensates, trapped in static potentials. Deviations from the 1D
behaviour in elongated and nearly spherical traps are treated e.g.
in~\cite{Muryshev1999,Feder2000,Muryshev2002}.

\section{Creation of dark solitons}
For the experimental investigation of the dark soliton velocity
field we first produce ground state Bose-Einstein condensates of
$^{87}$Rb as described in~\cite{Bongs1999a}.
The condensates typically contain $1.5\times 10^5$ atoms in the
($F$=2, $m_F$=+2) state, with less than $10\%$ of the atoms being
in the thermal cloud. The fundamental frequencies of our static
magnetic trap are $\omega_x=2\pi\times 14\,$Hz and
$\omega_{\perp}=2\pi\times 425\,$Hz along the axial and radial
directions. The condensates are cigar-shaped with the long axis
($x$-axis) oriented horizontally and an aspect ratio of
$\lambda\approx 30$, providing an essentially one dimensional
potential for the evolution of the dark soliton on the time scale
of the experiment.

\begin{figure}
   \begin{center}
   \epsfxsize=2in
   \epsfbox{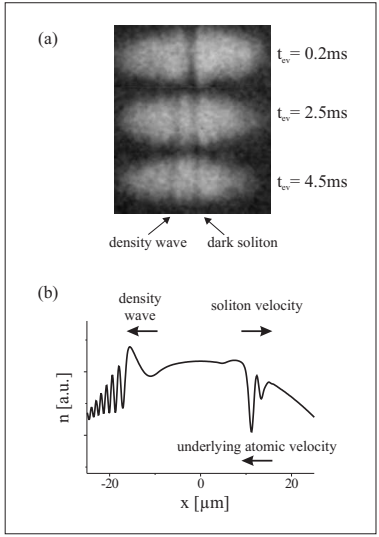}
   \end{center}
\vspace{0in} \caption {\small a) Absorption images of
Bose-Einstein condensates which evolved in the magnetic trap for a
time, $t_{ev}$ after phase imprinting. The images were taken after
an additional time-of-flight of 4$\, $ms, leading to density
minima for the dark soliton as well as for a density wave
\cite{Burger1999a}. b) Numerical simulation of the density
distribution of a dark soliton propagating in a BEC. The density
data shown was obtained for the same parameters as used for the
velocity field in Fig. \ref{fig6}.} \label{fig2}
\end{figure}

In our experimental setup we create dark solitons using a phase
imprinting method~\cite{Dobrek1999a,Burger1999a}. The phase
distribution of the condensate is changed by using the dipole
potential of a far detuned light field ($\lambda=$532\,nm,
$I\approx 20\,$W/mm$^2$), illuminating one half of the condensate
for a time $t_p$ on the order of $ 20\,\mu$s. Note that this time
has to be shorter than the correlation time of the condensate,
$t_c = \hbar /\mu$, in order to obtain pure phase imprinting,
where $\mu$ is the chemical potential and in our case $t_c\approx
55 \, \mu $s. This procedure creates a moving dark soliton and a
counterpropagating density wave, which become well separated after
some time, $t_{ev}$, of further nonlinear evolution in the trapped
Bose-Einstein condensate, as shown in Fig. \ref{fig2} a).

Figure~\ref{fig2} b) shows a typical density-distribution of a
dark soliton state created by phase imprinting after some
evolution time in a trapped BEC obtained by the numerical solution
of the GPE in 1D using the split operator
method~\cite{Burger1999a}. This simulation contains a finite
optical resolution for the phase imprinting field, leading to a
potential gradient with a width of $2\, \mu $m at the edge of the
illuminated area, as realized experimentally. Due to this gradient
a density wave and some higher order excitations moving to the
right are created in addition to the dark soliton. The dark
soliton and the density wave move in opposite directions, however
the dark soliton is much narrower due to its self stabilization in
the nonlinear medium. Note that the velocity field in the region
of the dark soliton has only components of negative velocities,
$v(x)<0$ for a positive soliton velocity $v_k>0$. This can be
intuitively understood by the fact that a movement of the density
minimum necessitates a transfer of matter in the opposite
direction.

\section{Bragg spectroscopy}
Bragg diffraction has been used as a tool to manipulate
Bose-Einstein condensates \cite{Kozuma1999a} in a number of
experiments. It was used to evaluate the velocity distribution of
a BEC within the magnetic trap and during time-of-flight
experiments \cite{Stenger1999b}. It has also served as a
beam-splitter in an interferometer \cite{Torii2000a}, for
interferometric studies of the phase evolution of BECs
\cite{Simsarian2000a} and even as an outcoupling mechanism for an
atom laser \cite{Hagley1999a}. In addition Bragg diffraction was
employed as a tool for the realization of exciting experiments on
non-linear effects in BECs \cite{Deng1999a} and their possible
application to create entanglement \cite{Vogels2002a}. However, a
complete review of the use of Bragg spectroscopy in experiments on
BEC is beyond the scope of this paper.

\begin{figure}
   \begin{center}
   \epsfxsize=3in
   \epsfbox{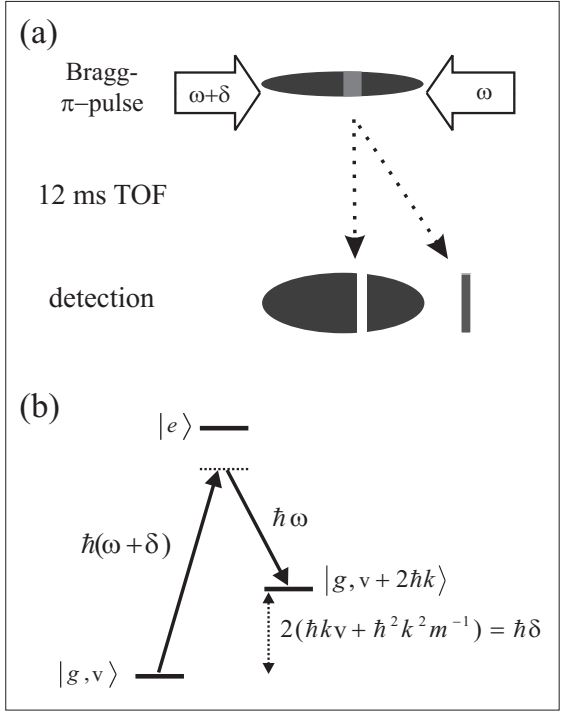}
   \end{center}
\vspace{0in} \caption {\small a) Schematic of position resolved
Bragg-spectroscopy of a dark soliton state in a BEC. b) Energy
level scheme for the Bragg transition between ground state levels
with velocities $v$ and $v+\frac{2\hbar k}{m}$ where $k$ is the
wave number of the laser field. The resonance condition for a
given frequency difference $\delta $ is only fulfilled for a
certain velocity $v$.} \label{fig3}
\end{figure}

Fig.~\ref{fig3} schematically shows the Bragg diffraction process
used in the experiments described here (a more detailed
explanation can be found in \cite{Kozuma1999a}). Two
counterpropagating laser beams induce a two photon transition
between two momentum states which differ by two photon momenta,
$2\hbar k$. Thus the atoms which have been in resonance with the
Bragg-transition will be separated from the remaining atoms by
$\Delta x=\frac{2\hbar kt_{\rm tof}}{m}$, where $t_{\rm tof}$ is
the time-of-flight after the Bragg pulse. Since the resonance
conditon depends on the initial velocity of the atom, a given
difference frequency of the laser beams addresses a specific
velocity class within the condensate. The width of this resonance
corresponds to the Fourier transform of the applied laser pulse
envelope. For $^{87}$Rb and counterpropagating laser beams the
velocity dependence of the resonance condition is given by
$\frac{\Delta v}{\delta /2\pi }\approx 0.39\frac{\rm mm/s}{\rm
kHz} $.

Experiments employing Bragg diffraction on BEC can be divided in
two general areas. By using short, intense pulses all velocity
classes within a BEC can be accessed and transferred into a moving
state. This process corresponds to a variable, coherent
beamsplitter for matterwaves. On the other hand, long pulses are
selective to the velocity in the BEC and distinct velocity classes
can be addressed. This process is used to examine the flux
underlying dark solitons in Bose-Einstein condensates. We show
that it is also possible to extract spatially resolved information
from these measurements, rather than only counting the outcoupled
number of atoms.

The experimental sequence starts with the creation of a
Bose-Einstein condensate in the magnetic trap followed by phase
imprinting for the creation of the soliton (as discussed above).
After phase imprinting the condensate evolves in the magnetic trap
for a time $t_{ev}$ which is varied in order to observe the
soliton excitation during formation and later propagation. The
Bragg scattering takes place after switching off the magnetic
trapping potential. For the experiments a free expansion time,
$t_{exp}=2\,$ms, is used after release from the trap in order to
reduce the atomic density before applying the Bragg pulse.
Collisions between atoms in different Bragg orders are thus
negligible. In order to suppress spontaneous processes the
frequencies of the Bragg lasers are detuned from the transition
$|F=2\rangle \rightarrow |F'\rangle$ by 6.8\,GHz. The Bragg pulse
is usually applied for a duration of $t_{Bragg}=250\,\mu$s, and a
time-of-flight of $t_{TOF}=13\,$ms after the pulse is used before
detection ($t_{Bragg}=500\,\mu$s and $t_{TOF}=11\,$ms for the BEC
spectroscopy in Fig. \ref{fig7}).

\section{Velocity selective spectroscopy}

The velocity selective Bragg scattering process described above is
a widely applicable tool to investigate the behaviour of
Bose-Einstein condensates. In the following we first observe the
formation process of dark solitons due to the phase imprinting
method. This is achieved by monitoring the spatial velocity almost
immediately after the phase has been imprinted. In a second set of
experiments the soliton is observed after a longer evolution time,
$t_{ev}$, in the magnetic trap, in order to investigate the
associated matter flux.

\subsection{Soliton formation}
\begin{figure}
   \begin{center}
   \epsfxsize=2in
   \epsfbox{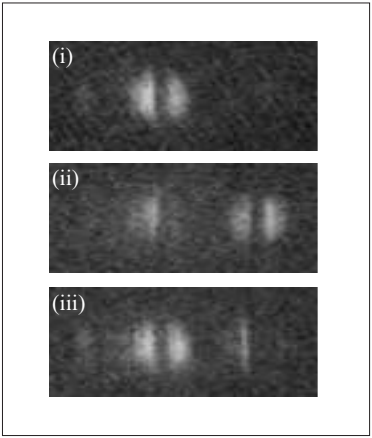}
   \end{center}
\vspace{0in} \caption {\small a) Absorption images for different
detunings $\delta$ of the Bragg-pulses: (i) off-resonant, (ii)
resonant with velocities $v=0$, (iii) resonant with velocities
$v<0$. These images were obtained for a short evolution time in
the magnetic trap of $t_{ev}=0.2\,$ms.} \label{fig4}
\end{figure}

Fig.\,\ref{fig4} shows absorption images of atomic clouds after
the above Bragg spectroscopy scheme has been applied with
$t_{ev}=0.2\,$ms and different resonance conditions. The upper
image (i) was obtained for an off-resonant Bragg pulse and thus
just corresponds to a time-of-flight image of the BEC after phase
imprinting. Image (ii) shows the case of a Bragg pulse resonant
with the resting atoms, which thus obtain a velocity, $2\hbar k$,
and appear to the right of their position in image (i) due to
their horizontal movement during the time-of-flight period of
$13\, $ms after phase imprinting. The moving atoms within the
cloud are not resonant with the Bragg pulse and remain on the
left. They correspond to the density wave and dark soliton. The
velocity of these atoms can be determined to be negative, as
expected, by scanning the Bragg frequency difference. The density
distributions corresponding to Bragg scattering resonant with
negative velocity atoms are shown in image (iii), with the resting
atoms on the left and the moving atoms on the right.

\begin{figure}
   \begin{center}
   \epsfxsize=3in
   \epsfbox{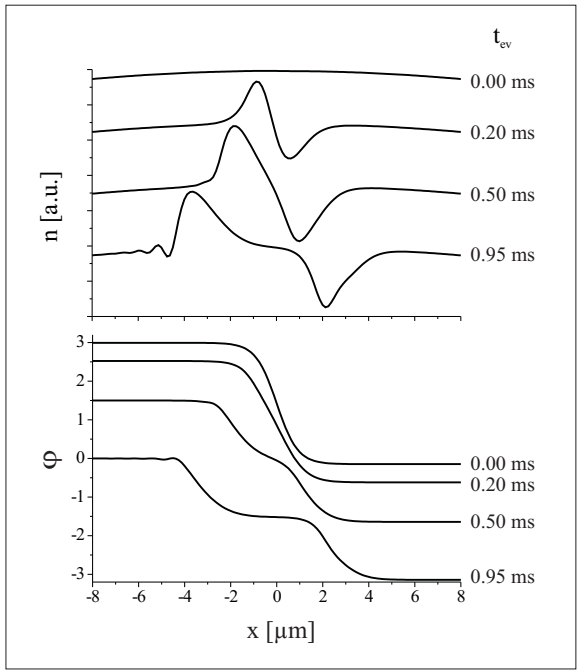}
   \end{center}
\vspace{0in} \caption {\small Density and phase of the condensate
wavefunction from numerical solutions of the Gross Pitaevskii
equation for short evolution times in the magnetic trap after
phase imprinting. For these simulations the width of the light
field edge was assumed to be $2\, \mu $m. These simulations show
that for the experimental parameters the soliton and density wave
are not yet distinguishable. Note that in this case the simulation
only shows the central part of the condensate and doesn't include
any time of flight.} \label{fig5}
\end{figure}

The simulations show that the velocity field in the Bose-Einstein
condensate after the short evolution time, $t_{ev}=0.2\, $ms
corresponds almost exclusively to the velocity field induced by
the phase imprinting pulse. Fig.\,\ref{fig5} shows calculated
density and phase distributions in the magnetic trap for various
times after the phase imprinting pulse. At $t_{ev}=0$ there is no
density change but only a phase gradient in the center of the
condensate, which corresponds to a velocity field. The associated
flow quickly leads to the formation of a density maximum, moving
to the left and a density minimum moving to the right. These two
start to separate after approximately $t_{ev}=0.5\, $ms with a
subsequent dispersive spreading of the density maximum and a
narrowing of the density minimum which leads to the dark soliton
state. Note that the initial phase gradient is split between the
density wave and the dark soliton since both are associated with
some flow of matter.

The evolution of the wavefunction in Fig.\,\ref{fig5} clearly
shows that the mechanism of phase imprinting with finite
resolution relies on the joint action of a phase shift and a local
removal of atoms due to a momentum transfer. The momentum imparted
on some parts of the wave function, assists in the later formation
of a dark soliton due to the removal of density from the region of
the soliton. The images in Fig.\,\ref{fig4} confirm these
numerical results. The imparted momentum can be explained
classically by the finite width of the edge of the phase
imprinting field, which produces a classical force due to the
gradient of the light induced AC-Stark shift potential.

\subsection{Flux associated with a dark soliton}\label{solitonflux}
\begin{figure}
   \begin{center}
   \epsfxsize=4in
   \epsfbox{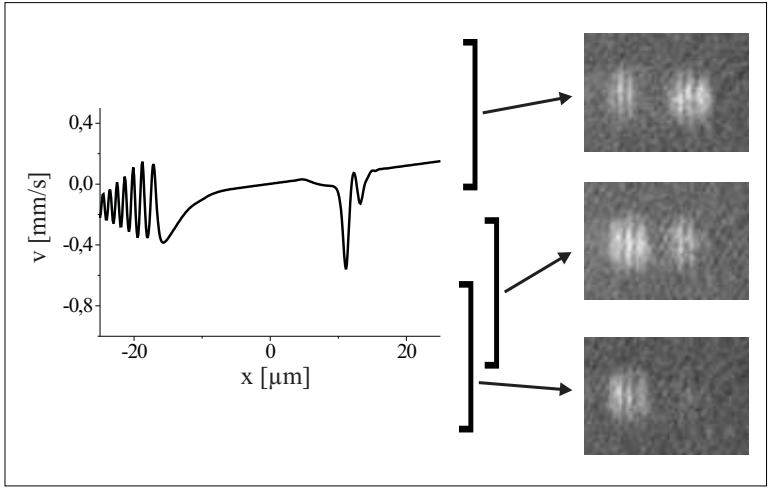}
   \end{center}
\vspace{0in} \caption {\small Left: velocity distribution of a
dark soliton state after phase imprinting with a phase difference
$\Delta \phi=\pi$, an evolution time in the magnetic trap
$t_{ev}=3.5\,$ms, and $t_{TOF}=2\,$ms of time of flight from a
numerical simulation of the Gross-Pitaevskii equation. Right:
corresponding Bragg spectroscopy images for different resonance
conditions of $v=+0.5\, $mm/s, $v=-0.7\, $mm/s and $v=-1.1\, $mm/s
(from top to bottom). The height of the vertical bars indicates
the width of the Bragg resonance, thus representing the velocity
classes, which will be scattered during the Bragg pulse.
 } \label{fig6}
\end{figure}

At short evolution times the dark soliton is masked by other
excitations induced by the phase imprinting pulse, but after a few
ms it becomes apparent due to its self stabilization and slow
propagation. An evolution time of $t_{ev}=3.5\, $ms is sufficient
to fully seperate the soliton and the density wave as shown by the
Bragg scattering experiments and the numerical simulation in Fig.
\ref{fig6}.

From the numerical simulation one expects to find local negative
velocity components around $v\approx -0.5\, $m/s at the positions
of the density wave as well as of the dark soliton. The top right
image shows nearly resting atoms scattered to the right, while two
distinct lines of moving atoms remain on the left: the atoms
associated with the dark soliton and the density wave. This
scattering behaviour is inverted for a negative velocity Bragg
resonance, now scattering the two lines to the right and leaving
the resting atoms on the left. No atom is effected by a Bragg
pulse resonant with larger negative velocities in the bottom
image, showing the limited range of the velocity field in the
matter wave. This is clear evidence of the matter wave flux
associated with a moving dark soliton.

\section{Mapping the velocity field}
In addition to the above examples of Bragg scattering for a few
distinct resonance velocities, which clearly show the signature of
dark soliton excitations, we use this method to fully map the
underlying velocity field.

\begin{figure}
   \begin{center}
   \epsfxsize=4in
   \epsfbox{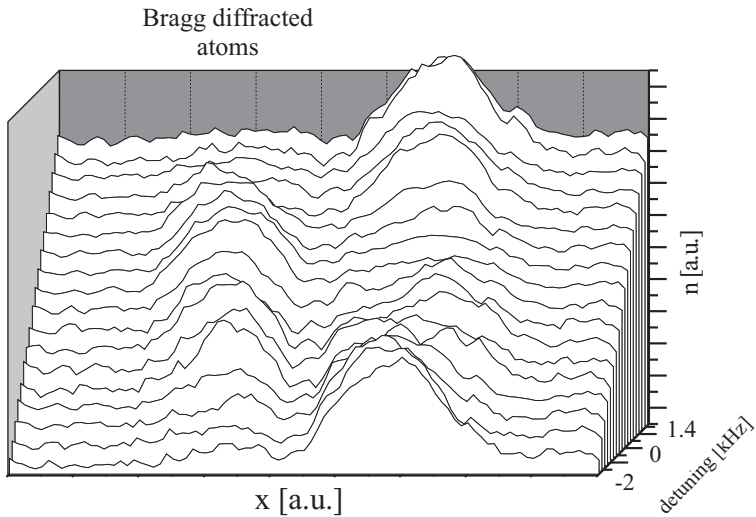}
   \end{center}
\vspace{0in} \caption {\small Vertically summed cuts through Bragg
spectroscopy images of a Bose-Einstein condensate without phase
imprinting. The Bragg pulse duration for this data was $500 \, \mu
$s. The Bragg resonance frequency is changed by 200$\, $Hz
(corresponding to $\Delta v\approx 0.08\, $mm/s) between
subsequent cuts. (Note that for technical reasons the position of
the scattered atoms is on the other side as for the other images.)
 } \label{fig7}
\end{figure}

We first explain and verify the scheme by mapping the spatial
velocity distribution of a pure Bose-Einstein condensate before
applying it to a dark soliton excitation. The starting point is a
systematic series of Bragg scattering data of a Bose-Einstein
condensate after 2$\, $ms of time of flight, as shown in Fig.
\ref{fig7}. By taking a vertical cut through the data sets in the
region of the scattered peak one obtains the amount of scattered
atoms at a fixed position as a function of Bragg detuning. The
maximum of this distribution then directly leads to the mean
velocity of the atoms scattered to this position. Their original
position in the condensate is then retrieved by accounting for
their time-of-flight displacement due to the imparted momentum and
their original velocity. Repeating this procedure for the position
range of the scattered peak this procedure directly gives the
velocity field of the condensate wavefunction. In addition to
analyzing the scattered peak, we cross check the velocity field
data by generating the same data from the inverse behaviour of the
non-scattered cloud.

\begin{figure}
   \begin{center}
   \epsfxsize=4in
   \epsfbox{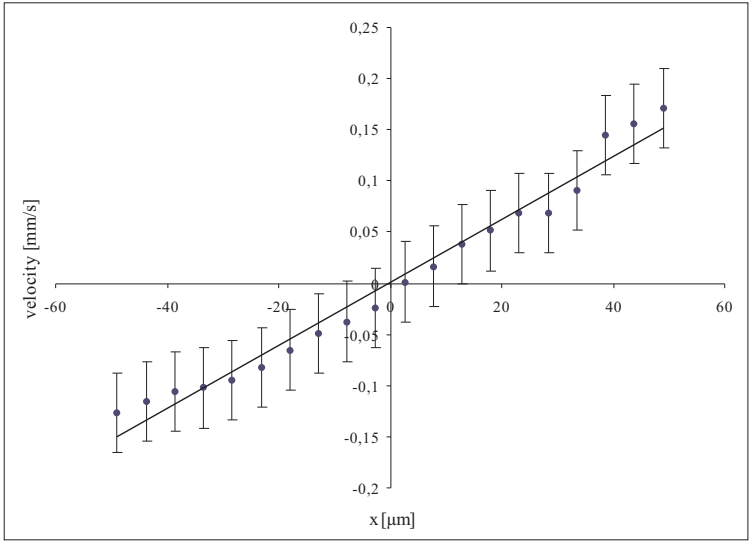}
   \end{center}
\vspace{0in} \caption {\small Velocity field of a freely falling
Bose-Einstein condensate as determined from the series of Bragg
scattering data in Fig. \ref{fig7}.
 } \label{fig8}
\end{figure}

Fig. \ref{fig8} shows the extracted velocity as a function of
position in the Bose condensate. The result is fitted with a
linear velocity distribution according to the expected behaviour
in the Thomas-Fermi limit. The good agreement confirms the
validity of this analysis.

\begin{figure}
   \begin{center}
   \epsfxsize=4in
   \epsfbox{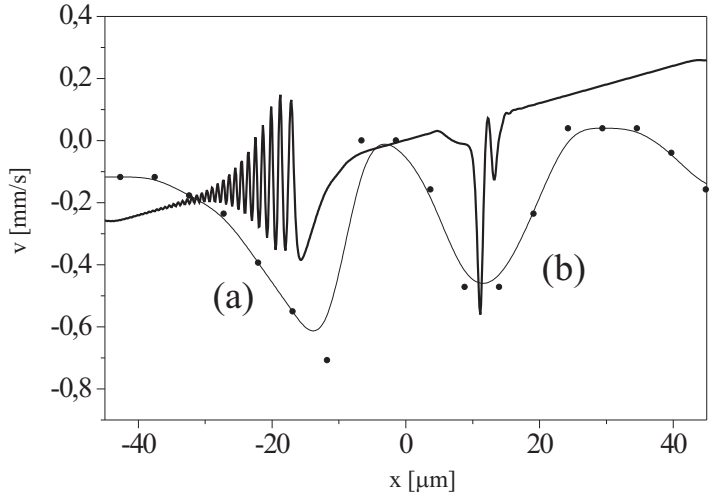}
   \end{center}
\vspace{0in} \caption {\small Velocity field underlying the dark
soliton excitation in section \ref{solitonflux} in comparison to
the numerical simulation from Fig. \ref{fig6}. The fit through the
experimental data is a spline fit to guide the eye. The negative
velocity peaks (a) and (b) correspond to the density wave and the
dark soliton respectively. Note that the averaging effect of the
experimental data is mainly due to the finite optical resolution
of $\approx 15\, \mu $m used for this set of measurements.
 } \label{fig9}
\end{figure}

By applying the above method to the dark soliton excitation
discussed in section \ref{solitonflux} we find the underlying
velocity field as shown in Fig. \ref{fig9}.

The comparison of the experimental data to the theoretical curve
shows good agreement within the finite experimental resolution,
which demonstrates the applicability of position resolved Bragg
spectroscopy to the investigation of excitations in Bose-Einstein
condensates.

\section{Conclusions}
In conclusion we have investigated the matter wave dynamics due to
phase imprinting and the matter wave flux underlying a moving dark
soliton. Both experiments and theoretical predictions clearly show
that the dark soliton movement is associated with a particle flux
in the opposite direction. It was demonstrated that Bragg
diffraction methods can be used to analyse the spatial phase and
the density distribution of dark soliton states in Bose-Einstein
condensates.

Future improvements of the tool of the position resolved Bragg
spectroscopy technique rely on the use of longer pulses and higher
resolution in the detection system. This should allow for
precision measurements on many kinds of excitations in
Bose-Einstein condensates. In this respect the investigation of
the interplay between the imprinted phase and the associated
imparted momentum promises important advances in the creation and
study of excitations in Bose-Einstein condensates.

\ack We gratefully acknowledge Anna Sanpera for help with the
numerical simulations and thank the DFG for support in the
Sonderforschungsbereich 407 as well as the European Union for
support in the RTN network ''Preparation and application of
quantum-degenerate cold atomic/molecular gases'' contract no:
HPRN-CT-2000-00125. \vspace{1cm}


\end{document}